\documentclass[11pt,titlepage,a4paper]{article}
\usepackage{bozhomac}	

\title{\bfseries	\vspace*{-1.6789in}
\vspace*{-8ex}
{
\begin{flushright}
	  \textbf{\large LANL xxx E-print archive No.~quant-ph/9803083}\\[2ex]
\end{flushright}
}
{\Large Fibre bundle formulation of\\
  	nonrelativistic quantum mechanics\\[1ex]
  \large 0. Preliminary considerations:\\[-0.3ex] Quantum mechanics
  	from a geometric-observer's viewpoint   }
}

\vspace{10pt}

\author{
Bozhidar Z. Iliev
\thanks{Department Mathematical Modeling,
Institute for Nuclear Research and \mbox{Nuclear} Energy,
Bulgarian Academy of Sciences,
Boul. Tzarigradsko chauss\'ee~72, 1784 Sofia, Bulgaria}
\thanks{E-mail address: bozho@inrne.acad.bg}
}

\newlength{\bo}\newlength{\ho}\newlength{\up}\newlength{\down}       
\newlength{\middle}                                                  
\newcommand{\bozho}{\leavevmode\hbox{\slshape\bfseries               
\settowidth{\bo}{BO}\settowidth{\ho}{HO}\settowidth{\middle}{/}
\settoheight{\up}{BOZHO}\settodepth{\down}{/}
\addtolength{\up}{+0.15\up}\addtolength{\bo}{+\middle}
\rule[\up]{\bo}{0.15ex}\hspace{-\bo}BO
\hspace{+0.09em}\raisebox{+0.17\up}{/}
\hspace{-0.20em}\raisebox{+0.71\up}{$\bullet$}
\hspace{-0.33em}\hspace{-1.14\middle}\raisebox{-0.4\up}{$\bullet$}
\hspace{-0.30em}\addtolength{\down}{-0.41\down}
\addtolength{\ho}{+1.5\middle}\rule[-\down]{\ho}{0.15ex}
\addtolength{\ho}{-\middle}\hspace{-\ho}\hspace{+0.18em}
\raisebox{+0.17\up}{HO}}}                                            
\newcommand{\BOZHO}
{\bozho$^{^{\text{\textregistered}\,} \text{\texttrademark} }$}      
\date{
\vspace{2.27ex}\ShortTitle{Bundle quantum mechanics: 0}	\\[0.27ex]
\vspace{5ex}
 Began:	14 February, 1995	\\
 Ended:	23 March, 1995	\\
 Revised: May 1996; April, September 1997	\\[1ex]
 Produced: \fbox{\today}	\\[1ex]
 Sent for submission to J. Phys. A: April 20, 1997 \\[1ex]
 LANL xxx archive server E-print No.~quant-ph/9803083 \\
 Published in
 J. Phys. A: Math. \& Gen., vol. 31, No.~4, \mbox{pp.~1297--1305, 1998.}\\
 \vspace{7ex}	{\Huge\bozho}	\\[48pt]
 \vspace{0.5ex} \Subject{Quantum mechanics; Differential geometry}\\[1.5ex]
\MSC[1991]{81P05, 81P99, 81Q99, 81S99}	\\[8pt]
\PACS[1996]{02.40.Ma, 04.60.-m, 03.65.Ca, 03.65.Bz} 	\\[8pt]
\KeyWords{Quantum mechanics; Geometrization of quantum mechanics}\\[8pt]
}

  \newcommand{\bs}[1]{\boldsymbol{#1}} 
\newcommand{\ih}{\mathrm{i}\hbar}

\begin{document}		

\renewcommand{\thefootnote}{\fnsymbol{footnote}} 
\maketitle			

\tableofcontents		


	\begin{abstract}

	We propose a version of the non-relativistic quantum mechanics in
which the pure states of a quantum system are described as sections of a
Hilbert (generally infinitely-dimensional) fibre bundle over the space-time.
There evolution is governed via (a kind of) a parallel transport in this
bundle.  Some problems concerning observables are considered. There are
derived the equations of motion for the state sections and observables. We
show that up to a constant the matrix  of the coefficients of the evolution
operator (transport) coincides with the matrix of the Hamiltonian of the
investigated quantum system.

	\end{abstract}

\section {\bfseries Introduction}
\label{I}
\setcounter{equation} {0}

	In conventional non-relativistic quantum mechanics a pure state
of some quantum system is described by state vector in a (generic
infinitely-dimensional) Hilbert space~\cite{Messiah-QM, Prugovecki-QMinHS}.
The time evolution of this vector is governed by the Schr\"odinger equation
but, for some purposes, it can also be represented (equivalently) via the
so-called evolution operator~\cite{Prugovecki-QMinHS}. In~\cite{Graudenz-94}
(see also~\cite{Graudenz-96} which is almost a review of~\cite{Graudenz-94},
but it also contains a new material) is suggested an interpretation of this
operator as a parallel transport in a (generic infinitely dimensional) vector
bundle over the space-time.

	Regardless that references~\cite{Graudenz-94,Graudenz-96} do not
meet any present-day mathematical standards of rigor, they do contain some
interesting ideas which we develop in the present work. On the one hand, we
accept the description of a quantum evolution as a (parallel) transport (of
sections) in a (Hilbert) fibre bundle over space-time. On the other, we
agree that quantities like the state vectors should generally explicitly
depend on the observer with respect to which they are referred, a fact which
is usually implicitly assumed.  Analogous feature can be found also in the
Prugove\v{c}ki's approach to the quantum theory
(see~\cite{Drechsler/Tuckey-96} for a selective summary) but we shall not
deal with it here. In the present work we apply these ideas to the
description of pure quantum states.

	This paper develops some aspects of the geometric approach to
non-relativistic quantum mechanics based on the Schr\"odinger equation.
We make an attempt to apply the theory of fibre bundles (and (linear)
transports on them) to quantum mechanics. In particular, we describe the time
evolution of pure quantum states, conventionally governed by the Schr\"odinger
equation, as a linear transport of sections (of a fibre bundle over the
space-time) along the trajectory (world line) of a given (local, i.e.
point-like) observer. It should be noted that this transport is not in the
`direction of time', it is along the observer's world line parameterized with
the (observer's proper) time. By means of the transport we transform section
values from one space-time point to another one. This `transportation' may be
towards the increasing as well as decreasing time values, which reflects the
fact that the Schr\"odinger equation (together with certain initial
condition(s)) predicts the wavefunction values in the future as well as in
the past.

	In Sect.~\ref{II} we briefly review the notion of a linear, in
particular parallel, transport along paths in vector bundles. In
Sect.~\ref{III} we make our basic assumptions concerning the geometry of
quantum mechanics. We suppose the (pure) states of a quantum system to be
described by sections of a Hilbert fibre bundle whose standard fibre is a
Hilbert space, isomorphic to the one of the conventional approach. This
bundle is assumed to be endowed with a (Hermitian fibre) metric by means of
which are determined the expectation (mean) values of the observables. The
time evolution of a system's state is governed by a (kind of a) parallel
transport found via the Schr\"odinger equation. This transport is
suppose to preserve the scalar products defined by the metric. In our
approach the observables are represented as bundle morphisms. In
Sect.~\ref{IV} we investigate certain consequences of the natural requirement
that the expectation values must be independent of the additional path via
which they are defined at points different from the one at which the
observer is situated.  Sect.~\ref{V} is devoted to the equations of motion
governing the time evolution of the state sections and observables. A
remarkable result here is that up to a constant the matrix of the Hamiltonian
coincides with the matrix of the coefficients of the evolution operator. In
this sense we can state that in our approach the Hamiltonian plays the role
of a gauge field. We close the paper with some remarks in
Sect.~\ref{conclusion}.

\section {Mathematical preliminaries}
\setcounter{equation} {0}
\label{II}

	In this section we recall some facts concerning linear transports
along paths in vector bundles \cite{f-LTP-general}.

	Let $(E,\pi,M)$ be a complex vector fibre bundle with base $M$, total
space $E$, and projection $\pi :E\to M$. The fibres $E_x:=\pi^{-1}(x)\subset
E$, $x\in M$, are isomorphic vector spaces, i.e. there exists a vector space
${\mathcal H}$ and isomorphisms $l_x$, $x\in M$  such that
$l_x:E_x\to {\mathcal H}$. We do not make any assumptions on the
dimensionality of $(E,\pi,M)$, i.e. ${\mathcal H}$  can has a finite as well
as infinite dimension. (Notice, the results
of~\cite{f-LTP-general,f-LTP-metrics} cited are valid also in the infinite
dimensional case regardless that they are proved there under the assumption
of finite dimensionality.)

	By $J$ and $\gamma:J\to M$ we denote a real interval and a path in
$M$, respectively.

	A \emph{${\mathbb C}$-linear transport (L-transport) along paths}
in $(E,\pi,M)$ is a map
$L:\gamma \mapsto L^{\gamma }$, where
$L^{\gamma }:(s,t)\mapsto L^{\gamma }_{s\to t}, \  s,t\in J$
is the (L-)transport along $\gamma$, and
$L^{\gamma }_{s\to t}:\pi ^{-1}(\gamma (s)) \to \pi ^{-1}(\gamma (t))$,
called (L-)transport along $\gamma$ from $s$ to $t$,
satisfies the equalities
\begin{eqnarray}
&
 L^{\gamma }_{t\to r}\circ L^{\gamma }_{s\to t}=L^{\gamma }_{s\to r},
\quad r,s,t\in J,
& \label{2.1} \\
&   L^{\gamma }_{s\to s}= {id}_{\pi ^{-1}(\gamma (s))}, \quad s\in J,
& \label{2.2} \\
&   L^{\gamma }_{s\to t}(\lambda u +\mu v)=
\lambda L^{\gamma }_{s\to t}u +
\mu L^{\gamma }_{s\to t}v, \quad \mu,\lambda\in {\mathbb C},
\quad u,v\in\pi ^{-1}(\gamma (s)).
& \ \ \ \label{2.3}
\end{eqnarray}
Here $id_N$ denotes the identity map of a set $N$.
The general form of $L^{\gamma }_{s\to t}$ is described by
\begin{equation}
 L^{\gamma }_{s\to t}={\left( F^{\gamma }_{t} \right)}^{-1}\circ
 F^{\gamma }_{s}, \quad s,t\in J       \label{2.31}
\end{equation}
with $F^{\gamma }_{s}:\pi ^{-1}(\gamma (s)) \to Q, \  s\in J$, being
one-to-one (linear) maps onto one and the same (complex) vector space Q.

	From~(\ref{2.1}) and~(\ref{2.2}) we see that
	\begin{equation}	\label{-1}
\left( L^{\gamma }_{s\to t}\right)^{-1}=L^{\gamma }_{t\to s}.
	\end{equation}

	According to~\cite[theorem~3.1]{f-TP-parallelT} the
set of (resp. linear) transports which are diffeomorphisms and satisfy the
locality and reparametrization conditions, i.e.  $ L^{\gamma }_{s\to t}\in
\mathrm{Diff}(\pi^{-1}(\gamma (s)),\pi^{-1}(\gamma (t)))$,
$ L^{\gamma |J^\prime }_{s\to t}=L^{\gamma}_{s\to t}$ for $s,t\in J^\prime $,
with $J^\prime $ being a subinterval of $J$, and
$ L^{\gamma\circ\tau}_{s\to t}=L^{\gamma}_{\tau(s)\to \tau(t)} $,
$s,t\in J''$ with $\tau$ being a 1:1 map of an ${\mathbb R}$-interval  $J''$
onto $J$, are in one-to-one correspondence with the (axiomatically defined
(resp. linear)) parallel transports (along curves). So, the usual parallel
transport along $\gamma$ from $\gamma (s)$ to $\gamma (t)$, assigned to a
linear connection, is a standard realization of the general (resp. linear)
transport $L^{\gamma }_{s\to t}$.

	Let $g$ be a (Hermitian) fibre metric on $(E,\pi,M)$,
i.e.~\cite{K&N-I} $g:x\mapsto g_x$ with $g_x:E_x\times E_x\to {\mathbb C}$,
$x\in M$, being nondegenerate Hermitian forms,
i.e.  $g_x$ are
Hermitian nondegenerate maps which are
${\mathbb C}$-linear in the second argument and ${\mathbb C}$-antilinear in
the first one.
%
A fibre metric $g$ and an L-transport $L$ are called \emph{consistent}
(resp.  along $\gamma$) if $L$ preserves the defined by $g$ scalar product,
i.e.~\cite{f-LTP-metrics}
\begin{equation}
g_{\gamma (s)}=g_{\gamma (t)}\circ\left( L^{\gamma }_{s\to t}\times
L^{\gamma }_{s\to t}\right),\quad s,t\in J \label{2.7}
\end{equation}
for all (resp. the given) $\gamma$.
In \cite{f-LTP-metrics} can be found different
results concerning this consistency.

	If $h:\mathcal{H}\times\mathcal{H}\to {\mathbb C}$ is a
Hermitian nondegenerate map which is
${\mathbb C}$-antilinear in the first argument and ${\mathbb C}$-linear in
the second one (a Hermitian metric
(scalar product) in ${\mathcal H}$), then, evidently, the map
$g:x\mapsto g_x:=h(l_x\cdot ,l_x\cdot ):E_x\times E_x\to {\mathbb C}$
is a fibre metric on $(E,\pi,M)$. Conversely, if $g$ is a fibre metric in
$(E,\pi,M)$ then, using the results from~\cite{f-LTP-metrics}, it can
easily be proved that the map
\(
h:=g_x(l^{-1}_x\cdot ,l^{-1}_x\cdot ):
\mathcal{H}\times\mathcal{H}\to {\mathbb C}
\)
is a Hermitian metric on
${\mathcal H}$ iff there is an L-transport along paths
consistent with $g$.%
\footnote{Notice that $h$ is always independent of $x\in M$.
	The transition $g\leftrightarrow h$ is similar to the one
in (gauge) gravitational theories, where (at a fixed point) one transforms a
general point-depending metric to the Minkowski one and vice versa, or,
equivalently, to the transition from a general basis to a local fierbein.}

	Let $\eta :J\times J^\prime\to M$ be a $C^2$ map. The curvature operator
${\mathcal{R}}^\eta (s,t):E_{\eta (s,t)}\to E_{\eta (s,t)}$ of the L-transport $L$
with respect to $\eta$ at $(s,t)\in J\times J^\prime$ is defined
by~\cite[equation~(3.1)]{f-LTP-Cur+Tor}.

	Let $\delta,\epsilon\in {\mathbb R}_+$ be such that
$(s+\delta,t+\epsilon)\in J\times J^\prime$ and $\lambda$ be the (oriented)
closed path defined as a product of the following paths:
$\sigma\mapsto\eta (s+\sigma,t)$ for $\sigma\in [0,\delta]$,
$\tau\mapsto\eta (s+\delta,t+\tau)$ for $\tau\in [0,\epsilon]$,
$\sigma\mapsto\eta (s+\delta-\sigma,t+\epsilon)$ for $\sigma\in [0,\delta]$,
and $\tau\mapsto\eta (s,t+\epsilon-\tau)$ for $\tau\in [0,\epsilon]$.
Hence $\lambda$ is a closed (oriented) loop connecting
the points: $\eta (s,t)$, $\eta (s+\delta,t)$,
$\eta (s+\delta,t+\epsilon)$, $\eta (s,t+\epsilon)$, and $\eta (s,t)$
in the written order.

	Supposing $L^{\gamma }_{s\to t}$ to have a $C^2$ dependence on $s$
(and thereof on $t$) and using~\cite[proposition 2.1]{f-LTP-general}, we
obtain, after some calculations, that the composition of the successive
L-transports of a vector at $\eta(s,t)$ along the paths forming $\lambda$  is
represented by an operator whose matrix has the following expansion
(see~\cite[Sect.~4]{f-LTP-Cur+Tor-prop})
	 \begin{equation}
\openone -
\delta\epsilon\bs{\mathcal{R}}^\eta(s,t)+
O(\delta^3)+O(\epsilon^3)+O(\delta^2\epsilon)+O(\epsilon^2\delta). \label{2.8}
	\end{equation}
\noindent
in some field of local bases. Here  $\openone$ is the unit matrix and
 $\bs{\mathcal{R}}^\eta(s,t)$ is the matrix corresponding to
 ${\mathcal{R}}^\eta(s,t)$.
If the L-transport along a product of paths is equal to the composition of
the L-transports along the corresponding paths of the product (in the
respective order), then this operator coincides with
the linear transport along $\lambda$.

\section {Basic differential-geometric assumptions}
\setcounter{equation} {0}
\label{III}

	The state of a quantum system will be described by a quantity
$\psi$ assumed to be a section of a vector bundle $(E,\pi,M)$ over
the space-time $M$:
\(
\psi\in\mathrm{Sec}(E,\pi,M) :=
\{\xi: \ \xi\colon M\to E,\quad\pi\circ\xi = {id}_M\}.
\)
The bundle $(E,\pi,M)$ is not supposed to be locally trivial. The typical
fibre ${\mathcal H}$ is supposed to be a Hilbert space, so such are all
(isomorphic to ${\mathcal H}$) fibers $E_x:=\pi^{-1}(x)$, $x\in M$.

	One can associate an L-transport along paths with the evolution
of any non-relativistic quantum system. For pure states this can be done
as follows (cf.~\cite{Graudenz-94}).
Let $\gamma :J\to M$
be the world line of an observer $B$. We interpret $t\in J$ as a proper time
(eigentime) of $B$. We suppose a quantum system to be described by $B$ at
$\gamma(t)\in M$, at the `moment' $t\in J$, by the state vector
$\psi _\gamma (t)\in E_{\gamma(t)}$, generally depending on $\gamma$
and $t$ separately; in particular it may depend only on $\gamma (t)$.
Let $B$ describe the evolution of the system with a Hamiltonian
$H_\gamma (t)$ through the Schr\"odinger equation, which in a matrix form
reads%
\footnote{%
In this work we present in a matrix form all relations containing derivatives.
In this way we avoid problems connected with the differentiation of fields of
objects defined (or acting) on $E$; e.g.
\( \partial \psi_\gamma(t)/\partial t \)
 is not (`well') defined at all. The invariant form of these relations will
be given elsewhere.%
} %
\begin{equation}	\label{2.4}
{d\over dt} \bs{\psi}_\gamma (t)=\bs{H}_\gamma (t)\bs{\psi}_\gamma(t).
\end{equation}
Here and from now on in our text we denote with bold symbols the
matrices  corresponding to vectors or operators in (a) given (field of)
bases (for details about infinite dimensional matrices - see,
e.g.,~\cite{Neumann-MFQM}).
We can write
\begin{equation}
\psi_\gamma (t)=U_\gamma (t,t_0)\psi_\gamma (t_0), \quad t,t_0\in J,
\label{2.5}
\end{equation}
where $t_0\in J$ is fixed and $U_\gamma (t,t_0)$ is
a linear operator, called \emph{evolution operator}, defined as the unique
solution of the initial-value problem~\cite{Prugovecki-QMinHS}
\begin{eqnarray}
& \ih {\partial\over {\partial t}}\bs{U}_\gamma (t,t_0)
		=\bs{H}_\gamma(t)\bs{U}_\gamma(t,t_0), & \label{2.6a}   \\
& U_\gamma (t_0,t_0)={id}_{E_{\gamma (t_0)}}.     & \label{2.6b}
\end{eqnarray}

	It is almost evident that
$ U_\gamma (t,t_0):E_{\gamma (t_0)}\to E_{\gamma (t)}$ is an L-transport
along $\gamma$ from $t_0$ to $t$, i.e.
$U:\gamma \mapsto U_\gamma :(t,t_0)\mapsto U_\gamma (t,t_0)$
is an L-transport along paths in $(E,\pi,M)$. Moreover, under certain natural
assumptions (cf.~\cite{Graudenz-94}), $U$ turns to be a (usual) parallel
transport.

	The fibre bundle $(E,\pi,M)$ is assumed to be endowed with two
structures: a linear transport along paths $L$, which is supposed to coincide
with the above-defined evolution operator $U$,%
\footnote{%
Later we preserve the notation $L$ as most of the results hold mathematically
for generic L-transport $L$, not only for the evolution operator $U$.%
}
and a consistent with it
Hermitian fibre metric $g$.  For brevity, as usual, we use the bracket
notation:
\begin{equation}
 \langle \psi (x)|\xi (x) \rangle_x :=g_x(\psi (x),\xi (x)),
\quad x\in M, \quad    \psi,\xi  \in\mathrm{Sec}(E,\pi,M). \label{3.1}
\end{equation}
So, now the consistency condition~(\ref{2.7}) reads
\begin{equation}	\label{3.2}
 \langle \psi (\gamma (s))|\xi (\gamma (s)) \rangle_{\gamma(s)} =
 \langle L^{\gamma }_{s\to t}\psi (\gamma (s))|
L^{\gamma }_{s\to t}\xi (\gamma (s)) \rangle_{\gamma(t)} .
\end{equation}

	Equation~(\ref{3.2}) restricts us
to consider only \emph{unitary} L-transports with respect to the metric.
In fact, if we define the
\emph{Hermitian conjugate} to $L^{\gamma }_{s\to t}$ transport,
\(
^{\dag } L^{\gamma }_{s\to t}:\pi ^{-1}(\gamma (s)) \to \pi ^{-1}(\gamma (t))
\)
by
\[
 \langle L^{\gamma }_{s\to t}\psi (\gamma (s))|
			\xi (\gamma (t)) \rangle_{\gamma(t)} =:
 \langle \psi (\gamma (s))|
	^{\dag } L^{\gamma }_{t\to s}\xi(\gamma (t)) \rangle_{\gamma(s)} ,
\]
then, due to~(\ref{-1}), we see~(\ref{3.2}) to be equivalent to
\(
 ^{\dag }L^{\gamma }_{s\to t} = L^{\gamma }_{t\to s}=
\left( L^{\gamma }_{s\to t}\right) ^{-1}.
\)%
\footnote{%
Dropping the arguments, if $\bs{U}$ and $\bs{G}$ are the matrices of the
transport and metric, respectively, the last equality is equivalent
to $^{\dag } \bs{U} = \bs{G}^{-1}\overline{\bs{U}}^\top \bs{G}$.
}

	Let ${\mathcal O}$ be the set of observables.
It's connection with the space-time is described by a map
$\varphi:{\mathcal O}\to
\mathrm{Morf}(E,\pi,M)$
assigning to $A\in{\mathcal O}$ a morphism $A_\varphi:E\to E$, i.e.
$\pi\circ A_\varphi=\pi$ (and hence $A_\varphi:E_x\to E_x$).

	The set of observers ${\mathbf{B}}$ consists of maps
$B_x:\mathrm{Sec}(E,\pi,M)\to E_x$, observers at $x$, assigning to any state
section $\psi$ a state vector at $x\in M$, i.e.
$B_x\colon\psi\mapsto\psi_B(x)$ 
.

	We define the \emph{expectation value} of
$A\in{\mathcal O}$ with respect to
$B_x$, when the system has a state section $\psi$, by
\begin{equation}
 \langle A \rangle _{B_x}:= {
 \langle \psi _B(x)|A_\varphi \psi_B(x) \rangle_x \over
{ \langle \psi_B(x)|\psi_B(x) \rangle_x } }.	\label{3.3}
\end{equation}

	The vector
$$
\psi^\gamma _{B,s,t}:=L^\gamma_{s\to t}\psi _{B}(\gamma(s))
$$
can be interpreted as a state vector of the quantum system at
$y=\gamma(t)$ `predicted' by an observer $B_x$ situated at $x=\gamma(s)$.
(Here $\gamma$ may not be the observer's world line.) By definition the
expectation value of $A\in{\mathcal O}$ at $y=\gamma(t)$ with respect to
$B_x$, $x=\gamma(s)$, along $\gamma$ is
\begin{equation}
 \langle A \rangle ^\gamma_{B,s,t}:=
{\langle \psi^\gamma_{B,s,t}|A_\varphi \psi^\gamma_{B,s,t}\rangle_{\gamma(t)}
\over{\langle \psi^\gamma_{B,s,t}|\psi^\gamma_{B,s,t}\rangle_{\gamma(t)} } }=
{ \langle \psi_B(x)|L^\gamma_{t\to s}\circ A_\varphi\circ L^\gamma_{s\to t}
\psi_B(x) \rangle_x \over{ \langle \psi_B(x)|\psi_B(x) \rangle_x}  },
\label{3.4}
\end{equation}
where~(\ref{3.2}) was used. Evidently, we have
$ \langle A \rangle ^\gamma_{B,s,s}= \langle  A \rangle _{B_{\gamma(s)}}$.

\section {Observables and the evolution operator}
\setcounter{equation} {0}
\label{IV}

	We assume the expectation value of $A\in{\mathcal O}$ at $y=\gamma(t)$
with respect to an observer $B_x$ to be independent of the path via which
it is determined, i.e. for $\beta:J^\prime\to M$ and
$\sigma,\tau\in J^\prime$, we demand
\begin{equation}
 \langle A \rangle ^\gamma_{B,s,t}=
 \langle A \rangle ^\beta_{B,\sigma,\tau}\ \mathrm{if} \
\beta(\sigma)=\gamma(s) \ \mathrm{and} \ \beta(\tau)=\gamma(t).  \label{4.1}
\end{equation}
This equality is a partial realization of the physical requirement that the
observed (expectation) values of the dynamical variables must be independent
of the way they are calculated.

	For a path $\alpha:J''\to M$ containing a closed loop at $x$, i.e.
$\alpha(s)=\alpha(t)=x$, for some $s,t\in J''$, this condition reduces to
$ \langle A \rangle ^\alpha_{B,s,t}= \langle A \rangle _{B_x}$
as we can choose $\beta$ to be
$\beta_\sigma:[\sigma,\sigma]\to\{x\}$. Using~(\ref{3.4}) we can rewrite
the last condition as
$ \langle \psi_B(x)|A_\varphi \psi_B(x) \rangle_x =   \langle \psi_B(x)|
L^\gamma_{t\to s}\circ A_\varphi\circ L^\gamma_{s\to t} \psi_B(x) \rangle_x$.
Admitting this equality to be valid for every $\psi_B(x)\in E_x$,
$x=\gamma(s)$, we get
\begin{equation}
[L^\alpha_{s\to t},A_\varphi]=0,
\quad \mathrm{for\ any}\ \alpha\ \mathrm{for\ which}\ \alpha(s)=\alpha(t),
\label{4.2}
\end{equation}
where $[\cdot,\cdot]$ denotes the commutator of the corresponding operators.
This result is a special case of the equation
\begin{equation}
[L^\gamma_{s\to t}\circ L_{\sigma\to\tau}^{\beta},A_\varphi]=0,
\quad \mathrm{for} \ \gamma(s)=\beta(\tau)\
\mathrm{and}\ \gamma(t)=\beta(\sigma)
\end{equation}
which is a corollary of~(\ref{3.4}) and~(\ref{4.1}).

	In particular, $A_\varphi$ commutes with the L-transport along any
closed path (loop) $\alpha$. Hence, if we choose $\alpha=\lambda$, with
$\lambda$ being the oriented closed path defined at the end of Sect.~\ref{II},
then for any L-transport satisfying the condition at the end of
Sect.~\ref{II}, we obtain
	 \begin{equation}
\left[ \mathcal{R}^\eta(s,t),A_\varphi\right]=0,	\label{4.3}
\end{equation}
where~(\ref{2.8}) was used, i.e. the curvature operator of the mentioned
linear transport commutes with all observables.  This is a necessary
condition for the validity of~(\ref{4.1}).

	As we have seen above, the map $L^\alpha_{s\to t}$ for
$\alpha(s)=\alpha(t)$
is independent of any local coordinates or trivialisations (if any),
it generally non-trivially transforms state onto state, and leaves the
observables invariant. Consequently it acts and can be considered as a
local symmetry transformation.

	The linear transport $L$ induces along $\gamma:J\to M$
the following transformation of an observable $A_\varphi$ , or,
more precisely, of
$\left.A_\varphi\right|_{E_{\gamma(t)}}$:
\begin{equation}
A_\varphi\mapsto A_\varphi^\gamma (s,t):=
L^\gamma_{t\to s}\circ A_\varphi\circ L^\gamma_{s\to t}
\ :E_{\gamma(s)}\to E_{\gamma(s)}.		\label{4.4}
\end{equation}
	In fact, $A_{\varphi}^{\gamma}(s,t)$  is the result of
`L-transportation' of $\left.A_\varphi\right|_{E_{\gamma(t)}}$ from  $t$ to
$s$ along $\gamma$. Rigorously speaking, the map
$\left.A_\varphi\right|_{E_{\gamma(t)}} \to A_{\varphi}^{\gamma}(s,t)$
is a linear transport along  $\gamma$ from $t$ to  $s$ in the fibre bundle of
bundle morphisms over $(E,\pi,M)$ (for details,
see~\cite[section~3]{f-TP-morphisms}).

	For the closed path $\lambda$
and special L-transports defined at the end of Sect.~\ref{II}
we can substitute~(\ref{2.8}) into~(\ref{4.4}). This gives
\[
\bs{A}^\lambda_\varphi (s,t) =
\left.\bs{A}_\varphi\right|_{E_{\lambda(s,t)}} +
\delta\epsilon\left[ \bs{\mathcal{R}}^\eta(s,t) , \bs{A}_\varphi \right] +
O((\delta,\epsilon)^3),
\]
where $O((\delta,\epsilon)^3)$ means third order quantities in
$\delta$ and $\epsilon$. Combining this with~(\ref{4.3}), we find
\begin{equation}
\bs{A}^\lambda_\varphi (s,t) =
\left.\bs{A}_\varphi\right|_{E_{\lambda(s,t)}} +
O((\delta,\epsilon)^3). 		\label{4.41}
\end{equation}

	Substituting equation~(\ref{4.4}) into~(\ref{3.4}), we get
\begin{equation}
 \langle A \rangle ^\gamma_{B,s,t}:=
{ \langle \psi_B(x)|A_\varphi^\gamma (s,t)  \psi_B(x) \rangle_x
\over{ \langle \psi_B(x)|\psi_B(x) \rangle_x }  }.  \label{4.42}
\end{equation}

	Due to~(\ref{3.2}), (\ref{3.4}) and~(\ref{4.1}), we, evidently, have
	\begin{eqnarray}
&  \langle \psi_B(x)|A_\varphi^\gamma (s,t) \psi_B(x) \rangle_x =
  \langle \psi^\gamma_{B,s,t}|A_\varphi \psi^\gamma_{B,s,t} \rangle_x =
& \nonumber  \\
& =  \langle \psi_B(x)|A_\varphi^\beta (s',t') \psi_B(x) \rangle_x , \qquad
\beta(s')=\gamma(s)=x,\ \beta(t')=\gamma(t).	       & \nonumber
	\end{eqnarray}

	If $r,r',s,t\in J$, then
$L^\gamma_{s\to t}=L^\gamma_{r'\to t}\circ
L^\gamma_{r\to r'}\circ L^\gamma_{s\to r}$ (see~(\ref{2.1})).
Inserting this equality into~(\ref{4.4}) and using~(\ref{-1})
we, after some algebra, obtain
\begin{equation}
A_\varphi^\gamma (r,s)\circ L^\gamma_{r'\to r} = L^\gamma_{r'\to r}\circ
A_\varphi^\gamma (r',t) \ :E_{\gamma(r')}\to E_{\gamma(r)}.  \label{4.5}
\end{equation}

	If $L^\gamma$ is a parallel transport along $\gamma$, then
putting here $\gamma=\beta^{-1}\alpha\beta$, where
$\beta:[a,b]\to M$, $\beta(a)=\gamma(r)=\gamma(r')$,
$\beta(b)=\gamma(s)=\gamma(t)$, and $\alpha:[a',b']\to M$,
$\alpha(a')=\alpha(b')$, we get
$$
\left[ L^\alpha_{a'\to b'},A^\beta_\varphi(a,b)\right]=0
$$
for every closed path $\alpha$ located at $y$ and any path $\beta$ containing
$y$ and $x$. However for general L-transports this equality may not hold.

	Let us assume that for the point $x\in M$ there is a neighborhood
$U\ni x$ such that $x$ can be connected by a path with any point from $U$.
Then there is a homotopy $\beta:U\times[0,1]\to M$
connecting $\chi_x:U\to{x}$ and
the inclusion map $\imath_U:U\to M,\ \imath_U(y)=y\in U$, i.e.
$\beta(\cdot,0):=\chi_x$ and
$\beta(\cdot,1):=\imath_U$.
Hence, the expectation value of $A\in{\mathcal O}$ at any $y\in U$
with respect to an observer $B_x$ is
\[
 \langle A \rangle ^{\beta(y,\cdot)}_{B}:=
 \langle A \rangle ^{\beta(y,\cdot)}_{B,0,1}=
 \frac{\langle \psi_B(x)|A^{\beta(y,\cdot)}_\varphi \psi_B(x) \rangle_x}
 {\langle \psi_B(x)|\psi_B(x) \rangle_x} ,
\]
where
\(
A^{\beta(y,\cdot)}_\varphi:=A^{\beta(y,\cdot)}_\varphi(0,1)=
L^{\beta(y,\cdot)}_{1\to 0}\circ A_\varphi\circ
L^{\beta(y,\cdot)}_{0\to 1}:E_x\to E_x.
\)

	Every intermediate point $\beta(y,\tau),\ \tau\in[0,1]$ is connected
with $x$ (besides via $\beta(y,\cdot)$) also by the path
$\beta_{y,\tau}:=\beta(y,\cdot)|_{[0,\tau]}:t\mapsto\beta(y,t)$ for
$t\in[0,\tau]$. We have
$$
 \langle A \rangle ^{\beta_{y,\tau}}_B:=A^{\beta_{y,\tau}}_{B,0,\tau}=
\frac{\langle \psi_B(x) | A^{\beta_{y,\tau}}_\varphi \psi_B(x) \rangle_x }
{\langle \psi_B(x) | \psi_B(x) \rangle_x} ,
$$
with
\begin{equation}
A^{\beta_{y,\tau}}_\varphi:=A^{\beta_{y,\tau}}_\varphi(0,\tau)=
L^{\beta_{y,\tau}}_{\tau\to 0}\circ A_\varphi\circ
L^{\beta_{y,\tau}}_{0\to \tau}:E_x\to E_x.        \label{4.6}
\end{equation}

	Let us assume that  the evolution of a quantum
system along $\beta_{y,\tau}$ is given by
$\psi_{\beta_{y,\tau}}(t)=L^{\beta_{y,\tau}}_{0\to t}\psi_{\beta_{y,\tau}}(0)$
through the Schr\"odinger equation~(\ref{2.4}), i.e. the L-transport
satisfies equation~(\ref{2.6a}):
$$
\ih {\partial \over{\partial t}}\bs{L}^{\beta_{y,\tau}}_{0\to t}=
\bs{H}_{\beta_{y,\tau}}(t) \bs{L}^{\beta_{y,\tau}}_{0\to t},
\ \  \mathrm{so \ that} \ \
\ih {\partial\over{\partial t}} \bs{L}^{\beta_{y,\tau}}_{t\to 0}=
 - \bs{L}^{\beta_{y,\tau}}_{t\to 0} \bs{H}_{\beta_{y,\tau}}(t).
$$

	Differentiating the matrix form of~(\ref{4.6}) with respect to $\tau$
and using these equalities, we get
\begin{equation}
\ih {\partial\over{\partial\tau}}\bs{A}^{\beta_{y,\tau}}_\varphi =
- \left[ \bs{H}^{\beta_{y,\tau}}_{\beta_{y,\tau}}(\tau) ,
	\bs{A}^{\beta_{y,\tau}}_\varphi\right]  \label{4.7}
\end{equation}
where
\(
H^{\beta_{y,\tau}}_{\beta_{y,\tau}}(\tau) :=
L_{\tau\to 0}^{\beta_{y,\tau}} \circ H_{\beta_{y,\tau}}(\tau) \circ
L_{0\to\tau}^{\beta_{y,\tau}}
\)
is the bundle morphism restricted on  $E_x$ corresponding to the Hamiltonian
$H_{\beta_{y,\tau}}(\tau)$ according to~(\ref{4.6}).

\section{Equations of motion}
\setcounter{equation} {0}
\label{V}

The Schr\"odinger equation~(\ref{2.4}) is an equation of
motion for the state vectors. Equation~(\ref{4.7}) plays the same role
with respect to observables. Below we consider briefly the analogues of these
equations in the theory considered here with linear transports.

	Let $B\in\mathbf{B}$ be an observer with a world line
$\gamma :J\to M$, i.e.
$B\colon x\mapsto B_x: \mathrm{Sec}(E,\pi,M)\to E_x,\ x=\gamma(s),\ s\in J$.
Let for a fixed
$s_0\in J$ the state vector of the quantum system be
$\psi_0:=\psi_\gamma(s_0)\in E_{\gamma(s_0)}$. We assume that along
$\gamma$ the state vector with respect to $B$ at $\gamma(s),\ s\in J$ is
obtained via some linear transport $L$ along paths, viz.
\begin{equation}
\psi_\gamma(s)=L^\gamma_{s_0\to s}\psi_0.	\label{5.1}
\end{equation}
This equation is our analogue of~(\ref{2.5}) and it plays the role of
the state vector (section) equation of motion.

	Let us define the matrix $\bs{\Gamma}_\gamma(s)$ of the
\emph{coefficients} of an L-transport by
\begin{equation}
\bs{\Gamma}_\gamma(s):=\left( {\partial\over{\partial s}}
\bs{L}^\gamma_{s\to t}\right) _{t=s}.			\label{5.2}
\end{equation}
Evidently (see~(\ref{2.31}))
\begin{equation}
\bs{\Gamma}_\gamma(s)= - \left( {\partial\over{\partial t}}
\bs{L}^\gamma_{s\to t}\right) _{t=s} =
\left( \bs{F}_{s}^{\gamma} \right)^{-1}
{\partial \bs{F}_{s}^{\gamma}\over{\partial s}}.  \label{5.3}
\end{equation}


	Now we shall prove that
\emph{%
up to a constant in our theory
$\bs{\Gamma}_\gamma(s)$  plays the role of a (matrix) Hamiltonian%
}
describing the
system's evolution through the Schr\"odinger-type equation. In fact,
from~(\ref{5.2}) and~(\ref{2.31}) we find
\begin{equation}
{\partial\over{\partial t}}\bs{L}^\gamma_{s\to t} =
 - \bs{\Gamma}_\gamma(t) \bs{L}^\gamma_{s\to t}.	\label{5.4}
\end{equation}
Combining this equation with~(\ref{5.1}), we confirm ourselves
that $\psi_\gamma(t)$ satisfies the Schr\"odinger equation~(\ref{2.4}) with
$\bs{H}_\gamma(t)=-\ih\bs{\Gamma}_\gamma(t)$, which proves our assertion.

	If system's evolution is described by a Hamiltonian
$H_\gamma(t)$ via~(\ref{2.4}), then our results hold for
$\bs{\Gamma}_\gamma(t) = -\bs{H}_\gamma(t)/\ih$.

	If $\bs{\Gamma}_\gamma(s)$ is a given matrix, then equation
(\ref{5.4}) with an initial condition~(\ref{2.2}) uniquely defines the
linear transport $L$.

	The matrix $\bs{\Gamma}_\gamma(t)$ can also be called a `gauge
matrix' as it defines the `extended (covariant) derivatives'. In fact,
recalling~\cite{f-LTP-general} that the deffe\-rentiation along paths
${\mathcal D}:\gamma\mapsto {\mathcal D}^\gamma$ defined by $L$ acts on a
$C^1$ section $\psi$ according to
\[
	\left({\mathcal D}^\gamma \psi \right)(s) :=
 	{\mathcal D}^\gamma_s \psi :=
\left. \left[ {\partial\over{\partial\epsilon}} \left(
L^\gamma_{{s+\epsilon}\to s}\psi(\gamma(s+\epsilon))
\right) \right] \right|_{\epsilon=0},		\label{5.5}
\]
we see that
${\mathcal D}^\gamma_s : \mathrm{Sec}^1(E,\pi,M)\to\pi^{-1}(\gamma(s)) $
and the matrix of the components of  ${\mathcal D}^\gamma_s \psi$ is
\(
{\partial\over{\partial s}}\bs{\psi}(s) +
		\bs{\Gamma}_\gamma(s) \bs{\psi}(s) .	\label{5.6}
\)

	The above discussion allows us to interpret the usual Hamiltonian as
a gauge operator, or, in some sense, as a `generalized affine connection'
along paths.

	Now to derive the generalization of~(\ref{4.7}) we have to
differentiate the matrix form of~(\ref{4.4}) with respect to $s$ and
use~(\ref{5.4}). Thus we get
\begin{equation}
{\partial\over{\partial s}}\bs{A}^\gamma_\varphi(s,t) = -\left[
\bs{\Gamma}_\gamma(s),\bs{A}^\gamma_\varphi(s,t)\right].	\label{5.7}
\end{equation}

	This is the equation of motion required for the observables.
In terms of the Hamiltonian, because of
\(
\bs{\Gamma}_\gamma(t) = - \bs{H}_\gamma(t)/\ih,
\)
it reads
\begin{equation}	\label{5.8}
\ih\frac{\partial}{\partial s} \bs{A}_{\varphi}^{\gamma}(s,t)  =
\left[
\bs{H}_\gamma(s) , \bs{A}_{\varphi}^{\gamma}(s,t)
\right].
\end{equation}

	Analogously, differentiating the matrix form of~(\ref{4.4}) with
respect to $t$, we find
\begin{equation}	\label{5.9}
\ih\frac{\partial}{\partial t} \bs{A}_{\varphi}^{\gamma}(s,t)  =
-
\left[
\bs{H}_\gamma(s,t) , \bs{A}_{\varphi}^{\gamma}(s,t)
\right]
\end{equation}
where
\(
\bs{\Gamma}_\gamma(t) = - \bs{H}_\gamma(t)/\ih
\)
was used and
\(
H_\gamma(s,t) :=
L_{t\to s}^{\gamma} \circ H_\gamma(t) \circ L_{s\to t}^{\gamma}
\)
is the morphism restricted on $E_{\gamma(s)}$ corresponding to the
Hamiltonian  $H_\gamma$. Equation~(\ref{5.9}) is an evident generalization
of~(\ref{4.7}) for arbitrary path $\gamma$.

\section {\bfseries Conclusion}
\label{conclusion}
\setcounter {equation} {0}

	The approach to non-relativistic quantum mechanics developed in this
paper is intended to bring it to the class of physical theories
mathematically based on the formalism of fibre bundles. At present level the
new approach is equivalent to the conventional one which will be proved in
detail elsewhere.

	The novel `bundle' treatment of old problems reveals new
possibilities for generalizations and interpretations (cf. the similar
advantages of the Prugove\v{c}ki's theory~\cite{Drechsler/Tuckey-96}). In
particular, it is likely that the bundle formalism in quantum theory will
be useful  for the unification of quantum mechanics and
gravitation. A reason for this hope is the fact that we have not used any
concrete model of space-time; it can be flat as well as curved one and,
generally, has to be determined by another theory like special or general
relativity.

	The fibre bundle formalism seems also applicable to relativistic
quantum theory and field theory which will be a subject of other works.
Since the purpose of the present paper is a geometric description of
the non-relativistic case, here we want only to make some comments on these
items.

	The fibre bundle approach to relativistic quantum mechanics,
generally, needs a different mathematical base than the one used in this
work. A typical example of this kind is a special-relativistic particle
described by the Klein-Gordon equation. An essential point here is that this
is a second-order partial differential equation with respect to time. This
implies that an initial value of the wavefunction is not sufficient for the
unique determination of its other values; for this one needs the initial
values of the wavefunction and its first time derivative. So, we can not
directly apply a `linear transportation' for getting wavefunction values
from one another (for details, see~\cite[Section~5]{f-LTP-appl}). A way to
overcome this problem is to consider a fibre bundle the elements of whose
fibres have two components formed from the wavefunction $\psi$ and its first
time derivative ${\partial\psi}/{\partial t}$, i.e. they are of the type
$(\psi,{\partial\psi}/{\partial t})^\top$.
Such two-component wavefunction satisfies a first order partial differential
equation with respect to time~\cite[ch.~XX, \S~5]{Messiah-QM}. This last
equation admits consideration analogous to the one of Schr\"odinger equation
presented in this paper. The above-pointed difficult does not arise for
particles described via the Dirac equation. In fact, since the Dirac equation
can be written as~\cite[ch.~XX, \S~6]{Messiah-QM}
 $\mathrm{i}\hbar{\partial\psi}/{\partial t}=H_D\psi$,
 $H_D$ being the Dirac Hamiltonian,
we can apply \textit{mutatis mutandis} the present investigation to Dirac
particles. For this purpose we have to replace the non-relativistic
Hamiltonian with the Dirac's one, the Hilbert space with the space of
4-spinors, etc.

	In connection with further applications of the bundle approach to the
quantum field theory, we notice the following. Since in this theory
the matter fields are represented by operators acting on (wave) functions
from some space, the matter fields in their bundle modification should be
described via morphisms of a suitable fibre bundle whose sections will
represent the (wave) function. We can also, equivalently, say that in this
way the matter fields would be sections of the fibre bundle of bundle
morphisms of the mentioned suitable bundle. An important point here is that
the matter fields are primary related to the bundle arising over the
space-time, not to the space-time itself to which are directly related other
structures, such as connections and the principle bundle over it.

 \bibliography{bozhopub,bozhoref}

%
\begin{thebibliography}{10}

\bibitem{Messiah-QM}
Messiah A.~M. L.
\newblock {\em Quantum mechanics}.
\newblock Interscience, New York, 1958.
\newblock (Vol.~I and vol.~II.).

\bibitem{Prugovecki-QMinHS}
Prugove\v{c}ki E.
\newblock {\em Quantum mechanics in Hilbert space}, volume~92 of {\em Pure and
  applied mathematics}.
\newblock Academic Press, New York-London, second edition, 1981.

\bibitem{Graudenz-94}
Graudenz D.
\newblock On the space-time geometry of quantum systems.
\newblock Preprint CERN-TH.7516/94, CERN, November 1994.

\bibitem{Graudenz-96}
Graudenz D.
\newblock The quantum gauge principle.
\newblock Preprint CERN-TH/96-107, CERN, April 1996.
\newblock (See also LANL xxx archive server No. hep-th/9604180).

\bibitem{Drechsler/Tuckey-96}
Drechsler W. and Tuckey~Ph. A.
\newblock On quantum and parallel transport in a {H}ilbert bundle over
  space-time.
\newblock {\em Classical and Quantum Gravity}, 13(4):611--632, April 1996.

\bibitem{f-LTP-general}
Iliev~B. Z.
\newblock Linear transports along paths in vector bundles. {I}.~{G}eneral
  theory.
\newblock JINR Communication E5-93-239, Dubna, 1993.

\bibitem{f-LTP-metrics}
Iliev~B. Z.
\newblock Linear transports along paths in vector bundles. {IV}.~{C}onsistency
  with bundle metrics.
\newblock JINR Communication E5-94-17, Dubna, 1994.

\bibitem{f-TP-parallelT}
Iliev~B. Z.
\newblock Transports along paths in fibre bundles. {II}.~{T}ies with the theory
  of connections and parallel transports.
\newblock JINR Communication E5-94-16, Dubna, 1994.

\bibitem{K&N-I}
Kobayashi S. and Nomizu K.
\newblock {\em Foundations of Differential Geometry}, volume~I.
\newblock Interscience Publishers, New York-London, 1963.

\bibitem{f-LTP-Cur+Tor}
Iliev~B. Z.
\newblock Linear transports along paths in vector bundles. {III}.~{C}urvature
  and torsion.
\newblock JINR Communication E5-93-261, Dubna, 1993.
\newblock (LANL xxx archive server \mbox{No.~dg-ga/9704004}).

\bibitem{f-LTP-Cur+Tor-prop}
Iliev~B. Z.
\newblock Linear transports along paths in vector bundles. {V}. {P}roperties of
  curvature and torsion.
\newblock JINR Communication E5-97-1, Dubna, 1997.

\bibitem{Neumann-MFQM}
von Neumann~J.
\newblock {\em Mathematical foundations of quantum mechanics}.
\newblock Princeton Univ. Press, Princeton, New Jersey, 1955.

\bibitem{f-TP-morphisms}
Iliev~B. Z.
\newblock Transports along paths in fibre bundles. {III}.~{C}onsistency with
  bundle morphisms.
\newblock JINR Communication E5-94-41, Dubna, 1994.

\bibitem{f-LTP-appl}
Iliev~B. Z.
\newblock Linear transports along paths in vector bundles. {II}.~{S}ome
  applications.
\newblock JINR Communication E5-93-260, Dubna, 1993.

\end{thebibliography}
%
 \bibliographystyle{unsrt}

\end{document}